\documentclass{article}
\usepackage{spconf,amsmath,epsfig}
\usepackage{graphicx}
\usepackage{booktabs} 
\usepackage{multirow}
\usepackage{cite}
\usepackage{float}
\usepackage{titlesec}
\usepackage{amsmath}
\usepackage{bm}
\usepackage{graphicx}
\usepackage{subfigure}
\usepackage{marvosym}
\usepackage{url}

\let\OLDthebibliography\thebibliography
\renewcommand\thebibliography[1]{
  \OLDthebibliography{#1}
  \setlength{\parskip}{0pt}
  \setlength{\itemsep}{0pt plus 0.3ex}
}

\begin{document}\sloppy


\title{Quantitative Evaluation of driver's situation awareness \\in virtual driving through Eye tracking analysis}

\name{Yunxiang Jiang$^{a}$, Qing Xu$^a$ \Letter, Kai Zhen$^{a,b}$, Shunbo Wang$^{a}$, Yu Chen$^c$}

\address{$^{a}$College of Intelligence and Computing, Tianjin University, China  \\ $^{b}$China Automotive Technology and Research Center Co., Ltd. \\ $^{c}$Technical College for the deaf, Tianjin university of technology, China\\qingxu@tju.edu.cn}

\maketitle

\begin{abstract}
In driving tasks, the driver's situation awareness of the surrounding scenario is crucial for safety driving. However, current methods of measuring situation awareness mostly rely on subjective questionnaires, which interrupt tasks and lack non-intrusive quantification. To address this issue, our study utilizes objective gaze motion data to provide an interference-free quantification method for situation awareness. Three quantitative scores are proposed to represent three different levels of awareness: perception, comprehension, and projection, and an overall score of situation awareness is also proposed based on above three scores. To validate our findings, we conducted experiments where subjects performed driving tasks in a virtual reality simulated environment. All the four proposed situation awareness scores have clearly shown a significant correlation with driving performance. The proposed not only illuminates a new path for understanding and evaluating the situation awareness but also offers a satisfying proxy for driving performance.\\
\end{abstract}

\section{Introduction}
\label{sec:intro}
Situation awareness (SA) is the ability of an individual to detect, comprehend, and predict their surrounding environment and the current situation. The concept was initially introduced by Endsley in 1995 in the military field \cite{Endsley_SA}. It has since been widely adopted in various other disciplines due to its universal applicability. Currently, it is gaining increasing attention from academia and industry, particularly in the fields of network security and autonomous driving\cite{SA_poplu}.

In the context of driving, situation awareness refers to a driver's capacity to detect, comprehend, and predict the road conditions, vehicles, pedestrians, and other traffic participants \cite{Endsley_SA}. This forms the basis for assessing the current traffic situation and making appropriate decisions. Effective situation awareness enables drivers to timely identify potential hazards and changes, allowing them to make accurate driving decisions and enhance overall driving safety \cite{SA_safdriv}. In addition, driver situation awareness is essential for fostering good cooperation between human drivers and autonomous driving AI in collaborative driving environments \cite{co_driving}.

The current methods for quantifying situation awareness primarily consist of questionnaires, many of which require interrupting the task or interfering with the driver. For example, one commonly used technique is the Situation Awareness Global Assessment Technique (SAGAT) \cite{Endsley_SA}. Consequently, these methods are challenging to apply in real-world driving scenarios.

For drivers with (corrected) normal vision, visual information input constitutes most sensory information, making it the primary way to obtain situation awareness\cite{nuclear_SA}\cite{flight_SA}. Building on this, we utilize gaze movement data collected by eye tracker (embedded in HTC VR glasses \cite{HTC_Vive}) to quantitatively evaluate drivers' situation awareness in a simulated driving. 

Our methods eliminate the need to interrupt the driver's continuous driving and avoid the corresponding additional cognitive load because we use physiological indicators (specifically, eye tracking metrics) instead of requiring the driver to answer questions. Besides, our methods solely rely on data calculations and entropic computations, which bypass the subjective processing of language and thinking, making them more objective. Also we applied the Principal component analysis (PCA) \cite{PCA_Pearson}\cite{PCA_Hotelling} to perform dimension reduction, increasing the effectiveness of data use with the maximum possible information.

The experimental results have clearly proved that our methodology is effective. The proposed situation awareness scores, especially the overall score, exhibit a correlation with driving performance of strong significance.

\section{RELATED WORKS}
During driving task, Situation awareness is an important measure for drivers, especially in the context of collaborative driving involving human drivers and autonomous driving systems. Determining whether the driver is capable of taking over the driving task, or whether the autonomous driving program is required to take over the driving task, is crucial to enable effective collaboration between the driver and the autonomous driving program. Starting from the seminal work in situation awareness, there have been various attempts to quantify this abstract concept. These attempts can be categorized into three types as follows.

The first type is based on the questionnaire. These methods quantify situation awareness by asking subjects about the situation of their surroundings. Examples of these methods include the Situation Awareness Global Assessment Technique (SAGAT) and the Situation Awareness Rating Technique (SART) \cite{SART}. Most of the Questionnaire methods require interruption, which is unrealistic during actual driving. The second type is relevant to the behavioral observation. These methods assess situation awareness by observing the subjects' operational behavior and decision-making responses to infer their level of situation awareness. The so called subject matter experts (SME) \cite{SME} is a typical example of this type. The behavior observation method relies on the subjective judgment of professionals and lacks objectivity. In addition, experts' participation in scenario is required, resulting in high labor costs and difficulties in quantifying situation awareness automatically. The third type is based on physiological measures, including eye tracking, EEG and others \cite{EEG}. These methods are directly based on objective physiological data, bypassing the processing of human language and cognition, as a result, can achieve a more objective and quantitative situation awareness effect. One attempt to quantify situation awareness using eye tracking metrics is the study conducted by Van de Merwe, van Dijk and Zon in their flight simulation experiments. They utilized gaze rate and dwell time as Level1 SAand used a kind of SGE as an indicator of new information acquisition activity Level3 SA \cite{flight_SA}.

The eye tracking entropy mentioned above is widely considered to be related to situation awareness and, it has been used in many studies as a method to quantify situation awareness \cite{entropy_SA_use}. In particular, commonly used eye tracking entropies indicators include gaze transition entropy (GTE) and stationary gaze entropy (SGE) \cite{GTE_SGE}, and previous studies have shown their relationship with situation awareness \cite{flight_SA} \cite{nuclear_SA}. GTE uses the probability transition between Markov states to describe the transition between gaze points, and uses the conditional entropy based on the transition probability to deliver the complexity of gaze transition with respect to gaze dispersion \cite{GTE_SGE}. SGE, in which the Shannon  entropy, based on an equilibrium probability distribution for Markov transition matrix, is used to describe the degree of complexity for spatial patterns of fixation dispersion \cite{GTE_SGE}. In this paper, we use the eye-tracking to quantify situation awareness and, gaze movement based entropic metrics have been exploited for this purpose.


\section{METHODOLOGY}

\subsection{Perception SA Score $SA_{L1}$}

We define a perception event as follows: each time the subject gazes at an object that is different from the previous object, it is regarded as a perception. The count of perception events is recorded as $C_{Dect}$, which represents the number of times that the subject perform perception event during driving.


Considering that subjects who can perform more perception events during a same driving task is considered to have better perception SA score, we use $\frac{C_{prcp}}{M_{total}}$, which is the frequency at which the subjects made the perception event per gaze movement, to reflect a component of the perception SA score of the subject. Here ${M_{total}}$ is the total number of gaze points during the driving process.

In addition, we deem that each perception event has a certain importance for driving. In a single perception, the importance of the perception can be deemed as the number of gaze points (denoted as $m_{sglDect}$). We use an average importance $\mu({m_{sglDect}})$ for all the perception events as the other component of the perception SA score.


The perception SA score $SA_{L1}$ is defined as follows:

\begin{equation}
SA_{L1} = \frac{C_{Dect}}{M_{total}} * \mu({m_{sglDect}}).
\label{L1SA}
\end{equation}

\noindent This $SA_{L1}$ score provides a comprehensive analysis of the frequency of perception and the level of importance of each perception. The higher the value indicates that the subject possesses a better perception and greater $SA_{L1}$, consequently leading to a possible improved driving performance.

\subsection{Comprehension SA Score $SA_{L2}$}
We consider that at the same time as each perception event performs, a sustaining comprehension process of the detected object follows. We characterize each comprehension process from two perspectives: different degrees of comprehension and importance.

In order to estimate the comprehension degree of each comprehension process, we weight each process using a bell-shaped standard Gaussian distribution $G(m_{sglDect})$, which rewards the processes that the gaze number is close to the average and penalizes processes that the gaze number is distant from the average. The reason why we use Gaussian distribution is because insufficient gaze is generally considered to be challenges in fully comprehending the object, whereas excessive gaze may suggest difficulties in comprehension or distractions. 


In addition, in order to estimate the importance degree of each comprehension process, we feature the process using the number of gaze points $m_{sglDect}$. This approach follows in the same thinking as the definition of $SA_{L1}$ above and will not be repeated here.

To summarize, based on the two perspectives mentioned above, the comprehension SA score $SA_{L2}$ is defined as 
\begin{equation}
SA_{L2} = \frac{\sum_{sglDect}^{C_{Dect}} m_{sglDect} * G(m_{sglDect})}{M_{total}}.
\label{L2SA}
\end{equation}

This score provides a comprehensive analysis of the degree of comprehension and the importance of the comprehension, allows for the quantification of the average comprehension level of the subject for each perception. Obviously a comprehension SA score indicates a driving performance of the subject.

\vspace{-0.3cm}
\subsection{Projection SA score $SA_{L3}$}

We deem that gaze projection relies more on past information rather than solely on the current motion of the object.
Given that the eye tracker embedded in the headset display has an eye movement acquisition interval of 11 ms, which is largely shorter the average person's reaction time of about 270 milliseconds \cite{reaction_time}. Therefore, the correlation between the gaze movement changes can be regarded as the projection of the future trend based on the past information.


In our consideration, the correlation between historical and current changes in eye movement can be used to quantify how large subjects do projections. The following treatments were performed to quantify the changes, both from the perspectives of direction and velocity, in eye movement at each gaze point during the driving process:
\vspace{-0.4cm}
\begin{figure}[H]
    \centering
    \includegraphics[width=0.35\textwidth]{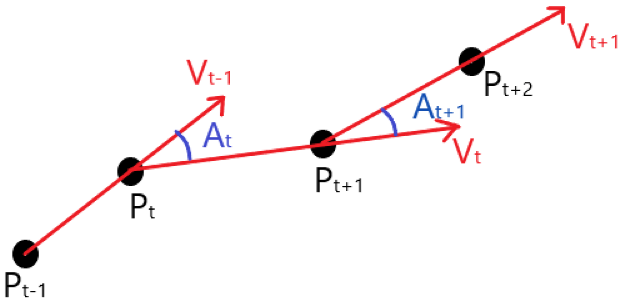}
    \caption{Illustration of the method used to quantify changes in gaze movements}
    \vspace{-0.3cm}
    \label{L3_graphic}
\end{figure} 
\vspace{-0.2cm}

\noindent
1) Forming a gaze vector $V_{t-1}$ by connecting the previous gaze point $P_{t-1}$ and the current gaze point $P_{t}$.

\noindent
2) Forming a gaze vector $V_{t}$ by connecting the current gaze point $P_{t}$ and the next gaze point $P_{t+1}$.


As shown in Fig.\ref{L3_graphic}, $A_t$ is the angle between two adjacent gaze motion vectors, $V_{t-1}$ and $V_{t}$, the calculation of this angle is used to quantify the extent of directional change in gaze motion. A specific calculation method we use is sine, $sin(A_t)$. Due to its sensitivity to small angles, the sine value of the angle between the two vectors is utilized to quantify the disparity in the direction of the two vectors at each given time. Simultaneously, the range of data values is constrained to $[0,1]$, thereby preventing any loss in accuracy resulting from the computation of inverse trigonometric functions. In addition, employing the vector's direction from the cross product to ascertain the magnitude of the sine value preserves a greater amount of directional information compared to relying solely on the angle value (as demonstrated in Fig.\ref{L3_graphic}, $sin(A_t)$ \textgreater 0 and $sin(A_{t+1})$ \textless 0).

The calculation of the length difference between the two gaze motion vectors, $V_{t-1}$ and $V_{t}$, is employed to quantify the extent of speed variation in the gaze motion. A specific calculation method taken is $-sign(V_t-V_{t-1})*log_{10}(\lvert V_t-V_{t-1} \lvert)$. Since the difference in gaze motion vector length is small and the length difference value is distributed around zero, we apply a logarithm base 10 transformation to the absolute length difference in order to amplify the small difference.


We make use of the powerful and entropic concept, mutual information \cite{mutual_infor}, for characterizing the correlation used in the definition of our projection SA score, to obtain $SA_{L3}^{dir}$ (direction perspective) and $SA_{L3}^{spd}$ (speed perspective) as 
\begin{equation}
SA_{L3}^{dir} = \sum_{t}(p(x_{t-1}^{dir},x_t^{dir})*log_2\frac{p(x_{t-1}^{dir},x_t^{dir})}{p(x_{t-1}^{dir})p(x_t^{dir})}
\label{L3SA_dir}
\end{equation}
\begin{center}
and
\end{center}
\begin{equation}
SA_{L3}^{spd} = \sum_{t}(p(x_{t-1}^{spd},x_t^{spd})*log_2\frac{p(x_{t-1}^{spd},x_t^{spd})}{p(x_{t-1}^{spd})p(x_t^{spd})},
\label{L3SA_spd}
\end{equation}
where $p$ represents the probability and $\{x_t\}$ denotes the sequence of speed difference or direction difference. In practice, we use the Gaussian KDE (kernel density estimation) method implemented in the python library 'scipy' \cite{scipy_KDE} for probability computations. $SA_{L3}^{dir}$ and $SA_{L3}^{spd}$ quantify the correlation between the direction and speed changes of three gaze vector pairs composed of every four adjacent gaze points, that is, the correlation between the change of gaze direction and the change of gaze speed. This is the core of our proposed projection SA score. The strength of the correlation between historical and future motion changes determines the subjects' inclination to predict future motion changes based on past data. A higher correlation indicates better predictive performance and a stronger L3 SA, and subjects exhibit improved driving performance. 



Principal Component Analysis (PCA) \cite{sklearn_PCA} was employed to obtain a single and complete projection SA score. The first principal component of PCA for situation awareness is obtained as the $SA_{L3}$ score 
\begin{equation}
SA_{L3}= PCA^{1stPC}(SA_{L3}^{dir},SA_{L3}^{spd}).
\label{L3SA}
\end{equation}

\subsection{Overall SA score $SA_{overall}$}
The PCA dimension reduction technique \cite{sklearn_PCA} is used to obtain a combined situation awareness, representing a so-called overall score of situation awareness. The overall SA score is defined as 
\begin{equation}
SA_{overall}= PCA^{1stPC}(SA_{L1},SA_{L2},SA_{L3}).
\label{overallSA}
\end{equation}
The first principal component of PCA of the three level SA scores is obtained as the overall SA score for situation awareness. This method captures the most significant variation in data, and retains as much effective information of the original three level SA scores as possible while reducing the data dimension.

\vspace{-0.5cm}
\section{EXPERIMENT}
\vspace{-0.5cm}

\subsection{Device}
To capture gaze movement data, the eye-tracking equipment embedded into the HTC Vive \cite{HTC_Vive} display is the 7INVENSUN Instrument aGlass DKII \cite{eyetracker}, which operates at a frequency of 90 Hz and provides gaze position accuracy of $0.5^{\circ}$. The Logitech G29 steering wheel serves as the driving device. Participants perceive ambient traffic and car engine sounds in the virtual environment (VE) through speakers. The experimental procedure can be observed on a desktop monitor, which displays the gaze movement and driving behaviors of the participants. Fig.\ref{device} provides an illustration of the headset with the embedded eye tracker.

\vspace{-0.5cm}
\subsection{Virtual reality environment}
\begin{figure}[H]
    \centering
    \begin{minipage}{0.24\textwidth}
        \centering
        \includegraphics[width=\textwidth]{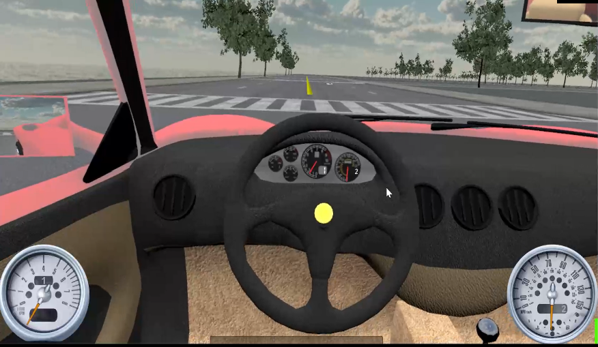} 
        \caption{Virtual environment}
        \label{environment}
        \centering
    \end{minipage}\hfill
    \begin{minipage}{0.19\textwidth}
        \includegraphics[width=\textwidth]{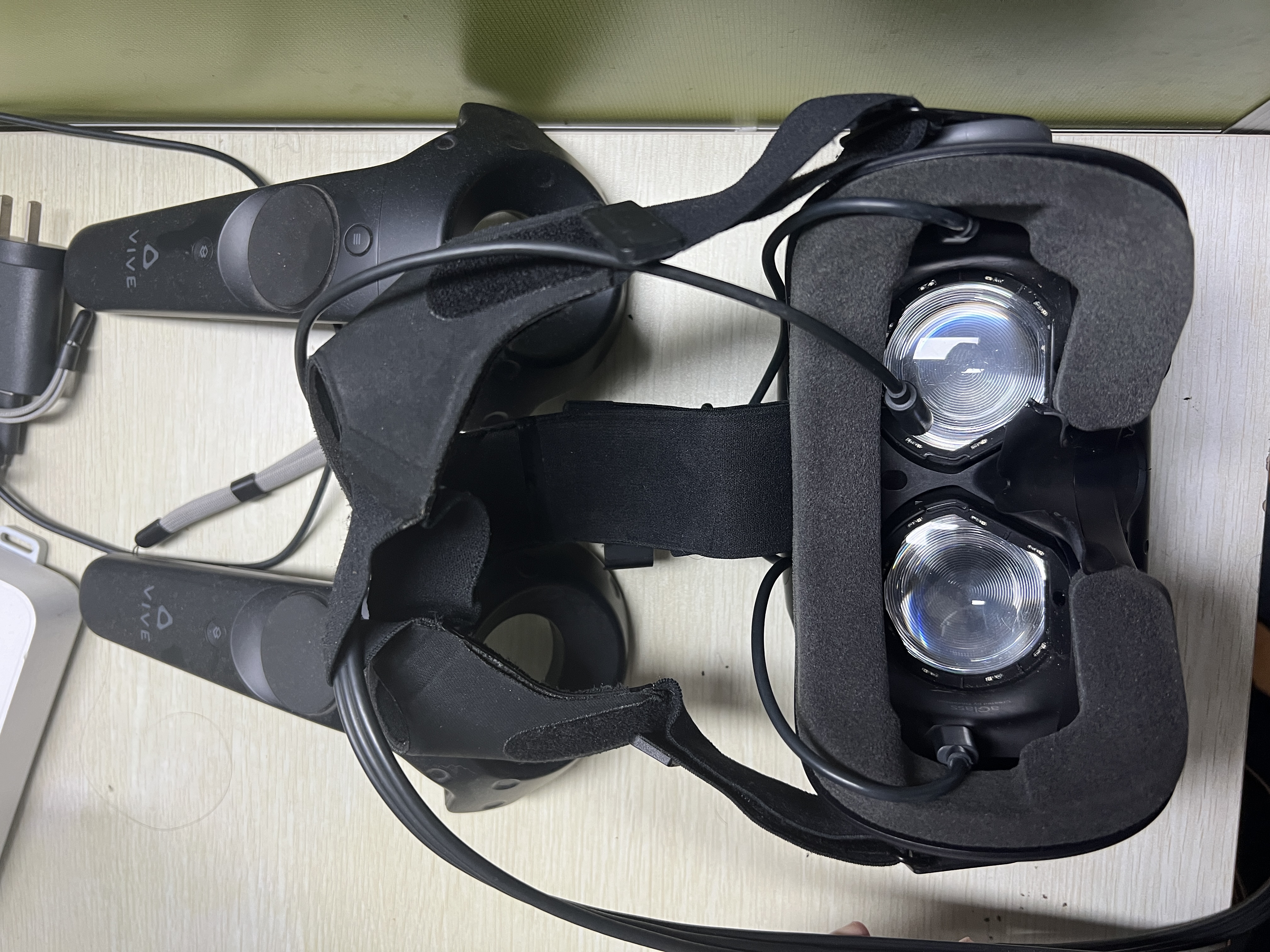}
        \caption{HTC headset}
        \label{device}
    \end{minipage}\hfill
    \vspace{0.1cm}
    \begin{minipage}{0.3\textwidth}
        \includegraphics[width=\textwidth]{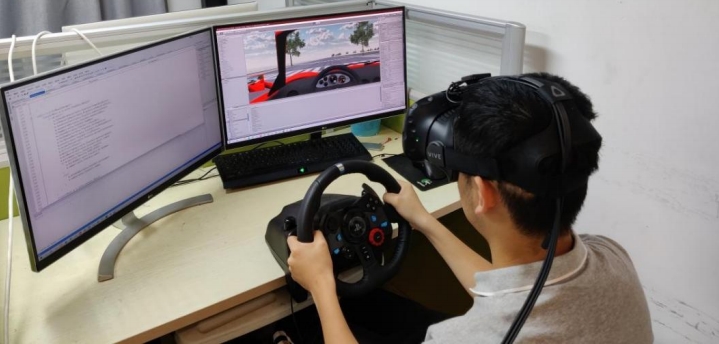}
        \caption{A subject is performing virtual driving task}
        \label{device}
    \end{minipage}
\end{figure}

\vspace{-0.3cm}
This study employs virtual reality as a tool to simulate actual driving scenarios, which has been extensively utilized in various similar studies. To facilitate the experimental analysis, a virtual environment is created, comprising of typical straight and curved roads along with buildings. An example of VR environment is presented in Fig.\ref{environment}.

In this study, we regard the gaze movement involved in senorita tasks as a task-oriented behavior, following the precedent set by numerous relevant works. Consequently, the virtual environment deliberately omits any visual distractors, such as pedestrians, traffic light signals.

\vspace{-0.6cm}
\subsection{Driving task}
\vspace{-0.2cm}
We designed driving tasks that both similar to normal driving behavior and reflect driving performance, the participants were asked to maintain a consistent speed of 40 km/h to ensure effective speed control. A lower mean acceleration indicates better driving performance, while a higher mean acceleration suggests poorer performance. We use the inverse of the average acceleration to represent driving performance.

\vspace{-0.6cm}
\subsection{Participants}
\vspace{-0.2cm}
Fourteen volunteers from our University participated in our psychophysical studies. The group consisted of 5 females and 9 males, with ages ranging from 21 to 25 (mean = 21.3, SD = 1.07). All the participants were recruited from our University to volunteer for the psychophysical studies and possess at least one and a half years of driving experience and hold a valid driver's license. Each participant has normal or corrected-to-normal visual acuity and exhibits normal color vision. Notably, none of the participants reported any adverse reactions to the virtual environment employed in our study.

\vspace{-0.6cm}
\subsection{Procedure}
\vspace{-0.2cm}
In this study, each test session is referred to as a trial, resulting in a total of $4*14 = 56$ valid trials, which can be considered satisfying the large-sample condition in classical statistics. \cite{samp_size} Throughout the test sessions, data on gaze movement and driving behaviors are recorded. Every participant undergoes four test sessions with identical task requirements and driving routes, with a one-week gap between each session. 

Before each test session, participants are provided with a preparation session to inform them about the objectives and steps involved in the psychological study.
\vspace{-0.3cm}

\vspace{-0,3cm}
\section{RESULTS}
\vspace{-0,3cm}
In the driving task, it is important for the situation awareness measures to accurately reflect the driving performance. To assess the effectiveness of our proposed scores in quantifying situation awareness, we conducted a study of correlation analysis using three commonly used correlation coefficients: Pearson linear correlation coefficient (PLCC), Spearman Rank correlation coefficient (SROCC), and Kendall Rank correlation coefficient (KROCC) \cite{cor_coif}. For comparison, we utilized entropy-based eye tracking indicators including GTE and SGE \cite{GTE_SGE} as well as direct indicators including gaze rate and dwell time \cite{dirct_eye_ind}, which have previously been used to quantify situation awareness \cite{nuclear_SA}\cite{flight_SA}. These situation awareness indicators solely based on eye movement data are specifically selected as comparison, because our proposed scores are based only on gaze data.

\vspace{-0.5cm}
\begin{table}[H]
\centering
\caption{CC between SA measures and driving performance}
\resizebox{0.5\textwidth}{!}{
\begin{tabular}{@{}l*{6}{c}@{}}
    \toprule
    \multirow{2}{*}{Measures} & \multicolumn{2}{c}{$Spearman$} & \multicolumn{2}{c}{$Kendall$} & \multicolumn{2}{c@{}}{$Pearson$} \\
    \cmidrule(lr){2-3}\cmidrule(lr){4-5}\cmidrule(l){6-7}
    & CC & p-value & CC & p-value & CC & p-value \\
    \midrule
            \textbf{L1 score} & 0.44 & $\textless10^{-3}$ & 0.31 & $\textless10^{-3}$ & 0.46 & $\textless10^{-3}$ \\
            \textbf{L2 score} & 0.36 & $\textless10^{-2}$ & 0.23 & $\textless10^{-2}$ & 0.28 & $\textless 0.05$ \\
            \textbf{L3 score} & 0.61 & $\textless10^{-6}$ & 0.44 &  $\textless10^{-5}$ & 0.60 &  $\textless10^{-6}$ \\
            \textbf{Overall score}  & 0.69 & $\textless10^{-8}$ & 0.51 &  $\textless10^{-7}$ & 0.67 &  $\textless10^{-7}$ \\
            GTE & 0.15 & $\textgreater 0.05$ & 0.04 & $\textgreater 0.05$ & 0.06 & $\textgreater 0.05$ \\
            SGE & 0.11 & $\textgreater 0.05$ & 0.05 & $\textgreater 0.05$ & 0.08 & $\textgreater 0.05$ \\
            Gaze rate & -0.01 & $\textgreater 0.05$ & 0.01 & $\textgreater 0.05$ & 0.02 & $\textgreater 0.05$ \\
            Dwell time & -0.13 & $\textgreater 0.05$ & -0.08 & $\textgreater 0.05$ & 0.04 & $\textgreater 0.05$ \\
    \bottomrule
\end{tabular}
}
\label{results_tab}

\end{table}

\begin{figure}[H]
    \centering
    \includegraphics[width=0.4\textwidth]{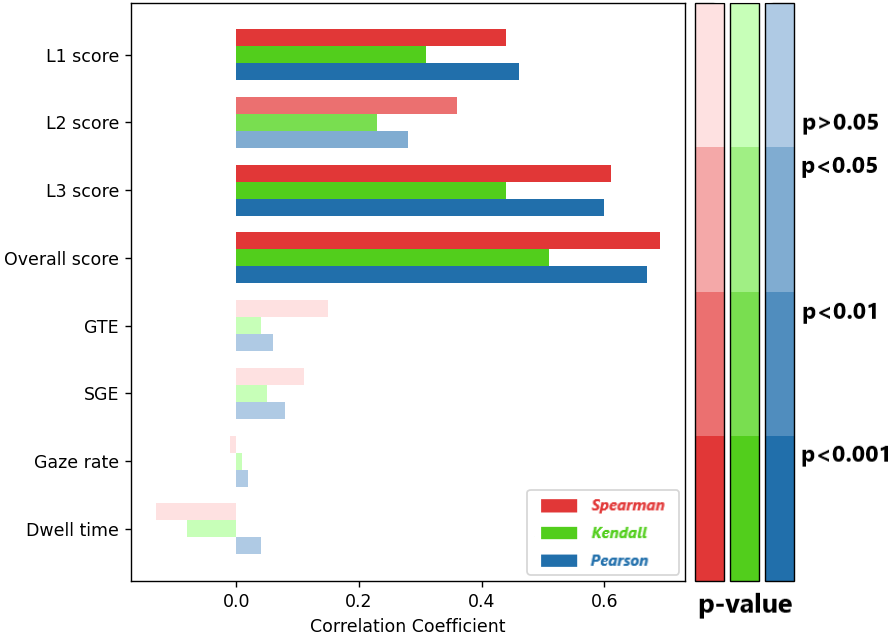}
    \caption{CC between SA measures and driving performance}
    \label{cola_result}
\end{figure} 
\vspace{-10pt}

The correlation results, as presented in Table.\ref{results_tab} and Fig.\ref{cola_result} respectively, demonstrate a significant correlation between our proposed scores and driving performance. On the other hand, direct (Gaze rate and Dwell time) and entropy-based eye tracking indicators such as GTE and SGE did not show a significant correlated relationship with driving performance. These findings suggest that our newly proposed situation awareness scores can be used as a reliable and proxy metric of driving performance and situation awareness.

\vspace{-0.3cm}
\begin{figure}[H]
    \centering
    \includegraphics[width=0.4\textwidth]{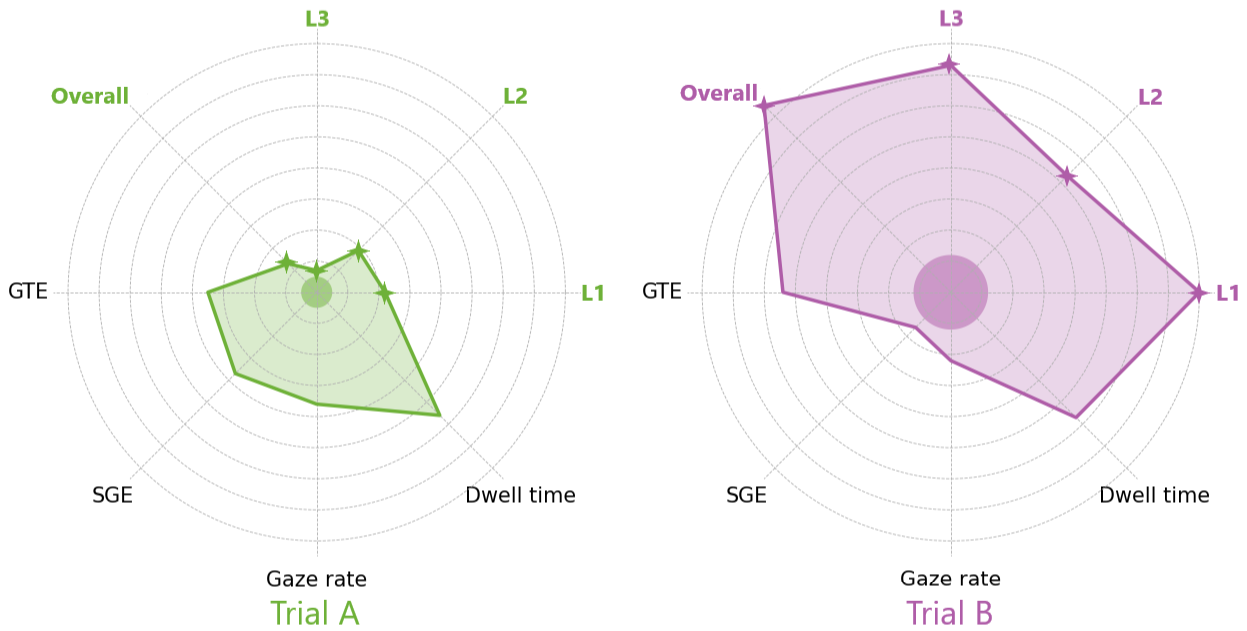}
    \caption{Radar plot of two representative examples}
    \label{radar}
\end{figure} 
\vspace{-0.3cm}

We present two representative examples Trials A and B in Fig.\ref{radar}, from the third and first test sessions conducted by the 12th and 7th participants, depicted in the radar plot. The size of the central circle indicates the driving performance of subject , as shown in this figure, the driving performance of Trial B (right) is largely better than that of Trial A (left). Each axis represents the corresponding measures, We highlight the four scores we have proposed in this paper. 

The radar chart demonstrates that all the four proposed scores for Trial B are larger than those for Trial A, coinciding with the driving performance of the two trials, while the other indicators under comparison do not show such pattern. This clearly indicates that our proposed scores effectively reflect the situation awareness and driving performance.

\vspace{-0.5cm}
\section{DISCUSSIONS}
\vspace{-0.2cm}
This study presents a novel behaviometric approach that utilizes eye tracking data to assess and comprehend the situation awareness of subjects while driving. To validate our findings, participants were instructed to perform in a virtual driving. The statistical analysis reveals a strong correlation between the proposed situation awareness scores and the performance of driving task.

In this study, Pearson, Spearman, and Kendall's coefficients were utilized for conducting correlation analysis. We applied the PCA to the three level situation awareness scores and take the first principal component as the overall situation awareness score. This overall score is essentially a linear combination of the three level ratings, which synthetically considers the participants' performance in detecting, comprehending, and predicting, while preserving as much  information as possible. The findings point out that the proposed three level and overall situation awareness scores exhibit a significant correlation with driving performance. The p-values obtained were generally below 0.01 or even 0.001, suggesting a statistical significance. The correlation coefficients (CC) ranged from 0.2 to 0.7, basically indicating a moderate degree of correlation \cite{cor_coif}. Notably the L3 and overall scores achieve a very significant correlation with strong significance.

In contrast, commonly used eye tracking entropy indicators such as GTE and SGE \cite{nuclear_SA}, and direct eye tracking indicators like gaze rate and dwell time \cite{flight_SA} do not exhibit a significant correlation with driving performance. We deem that the failure in these attempts may be attributed to that the definition of these indicators does not closely align with the concept of perception, comprehension, and projection of situation awareness.

Compared to the traditional questionnaire SA quantification method, our proposed method based on eye movement data eliminates the need for drivers to answer questions, thus avoiding the interruption of the driving task or the interference to the driving task. Additionally, our quantitative method relies solely on data calculations, bypassing subjective language and thought processing and making it more objective.

\vspace{-0.5cm}
\section{CONCLUSIONS}
\vspace{-0.2cm}
This paper contributes to a good understanding of the driver's situation awareness in virtual driving. Based on the objective gaze movement data, a methodology is proposed to objectively and quantitatively evaluate situation awareness. An overall score of situation awareness is also established. What we have achieved provides a non-intrusive way to evaluate a driver's situation awareness and serves as a reliable proxy for objectively assessing the performance of a driving task.

Basically, the current study employs the statistics methodology for situation awareness. For future, temporal dynamics of situation awareness could be investigated, and this could provide the possibility to study the causal relationship between the different levels of situational awareness, leading to a deeper understanding of situation awareness. Furthermore, in practical applications, this perspective of temporal dynamics could potentially offer industry solutions for quantitatively evaluating real-time situation awareness.

\vspace{-0.5cm}


\bibliographystyle{IEEEbib}
\bibliography{Citing}

\end{document}